\documentclass[superscriptaddress,prb,twocolumn,aps,showpacs,floatfix]{revtex4}
                 
\usepackage{graphicx}
\usepackage{rotating}
\usepackage{amsmath}
\usepackage{bbm}  
\usepackage{color}
    
\newcommand{\be}{\begin{equation}}
\newcommand{\ee}{\end{equation}}
\newcommand{\bea}{\begin{eqnarray}}
\newcommand{\eea}{\end{eqnarray}}

\renewcommand{\v}[1]{{\bf #1}}

\begin{document}

\title{Exchange-correlation orbital functionals in current-density-functional
theory: Application to a quantum dot in magnetic fields}

\date{\today}

\author{N. Helbig}
\affiliation{Unit\'e de Physico-Chimie et de Physique des Mat\'eriaux, Universit\'e Catholique de Louvain, B-1348 Louvain-la-Neuve, Belgium}
\affiliation{Institut f\"ur Theoretische Physik, Freie Universit\"at Berlin, Arnimallee 14, D-14195 Berlin, Germany}
\affiliation{European Theoretical Spectroscopy Facility (ETSF)}
\author{S. Kurth}
\affiliation{Institut f\"ur Theoretische Physik, Freie Universit\"at Berlin, Arnimallee 14, D-14195 Berlin, Germany}
\affiliation{European Theoretical Spectroscopy Facility (ETSF)}
\author{S. Pittalis}
\affiliation{Institut f\"ur Theoretische Physik, Freie Universit\"at Berlin, Arnimallee 14, D-14195 Berlin, Germany}
\affiliation{European Theoretical Spectroscopy Facility (ETSF)}
\author{E. R\"as\"anen}
\affiliation{Institut f\"ur Theoretische Physik, Freie Universit\"at Berlin, Arnimallee 14, D-14195 Berlin, Germany}
\affiliation{European Theoretical Spectroscopy Facility (ETSF)}
\author{E. K. U. Gross}
\affiliation{Institut f\"ur Theoretische Physik, Freie Universit\"at Berlin, Arnimallee 14, D-14195 Berlin, Germany}
\affiliation{European Theoretical Spectroscopy Facility (ETSF)}

\begin{abstract}
The description of interacting many-electron systems in  external magnetic
fields is considered in the framework  of the optimized effective potential
method extended to current-spin-density functional theory.  As a case study, a
two-dimensional quantum dot in external  magnetic fields is investigated.
Excellent agreement with quantum  Monte Carlo results is obtained when
self-interaction corrected  correlation energies from the standard local
spin-density approximation are  added to exact-exchange results. Full
self-consistency within the complete current-spin-density-functional framework 
is found to be of minor importance.
\end{abstract}


\maketitle

\section{Introduction}

Since its introduction in 1964, density-functional 
theory (DFT) \cite{HK1964,KS1965} has become a 
standard tool to calculate the electronic  structure of
atoms, molecules, and solids from first principles.  
Early on, the original DFT formulation has been 
extended to the case of  spin-polarized systems
\cite{BH1972} which also provides a description 
of many-electron systems in an external magnetic field. 
However, in this spin-DFT (SDFT) framework the
magnetic field only couples to the spin  but not 
to the orbital  degrees of freedom, i.e., the coupling 
of the  electronic momenta to the vector potential
associated with the external  magnetic field is 
not taken into account. A proper treatment of this 
coupling requires extension to 
current-spin-density-functional theory 
(CSDFT)~\cite{VR1987,VR1988} in terms of three basic 
variables: the electron density $n(\v r)$, the spin 
magnetization density $\v m(\v r)$, and the paramagnetic 
current density $\v j_p(\v r)$. These  densities are 
conjugate variables to the electrostatic potential, the
magnetic field, and the vector potential, respectively. 

In order to be applicable in practice, DFT of any flavor 
requires an approximation to the exchange-correlation (xc) 
energy functional. The use of the local-vorticity 
approximation,~\cite{VR1987, VR1988} which is an extension of
the local spin-density approximation (LSDA), is problematic
in CSDFT: the xc energy per particle of a uniform electron 
gas exhibits derivative discontinuities whenever a Landau level
is depopulated in an increasing external magnetic 
field. This leads to discontinuities in the corresponding xc
potential.~\cite{SV1993} These discontinuities then incorrectly 
appear when the local values of the {\em inhomogeneous} 
density and vorticity coincide with the corresponding values 
of the {\em homogeneous} electron gas. A popular way to
circumvent this problem is to use functionals which 
interpolate between the limits of weak and high magnetic 
fields.~\cite{WR2004b,SRSHPN2003} 

Explicitly orbital-dependent 
functionals, which are successfully used in DFT and 
collinear SDFT,~\cite{GKKG2000, KK2008} are natural candidates to 
approximate the xc energy in CSDFT for two reasons: first, 
they are constructed without recourse to the model of the 
uniform electron gas and second, they are ideally suited 
to describe orbital effects such as the filling
of Landau levels. In this way, the problem inherent in 
any uniform-gas-derived functional for CSDFT is avoided 
in a natural way. 

The use of orbital functionals  requires the so-called 
optimized effective potential (OEP) method~\cite{TS1976} 
to calculate the effective potentials. 
The OEP formalism has been recently generalized to
non-collinear SDFT~\cite{SDDHKGSN2006} as well 
as to CSDFT.~\cite{PKHG2006} In addition, a larger
set of basic densities has been considered 
in order to include the spin-orbit coupling.\cite{B2003,RG2006} 
Recent applications of the OEP method for atoms~\cite{PKHG2006} and
periodic systems~\cite{SPKSDG07} have indicated that
the difference between exact-exchange calculations 
carried out fully self-consistently within CSDFT 
or SDFT, respectively, is only minor. These works have also 
indicated that the inclusion of correlation energies is of particular 
importance when dealing with current-carrying states. 

In this work we consider the OEP formalism within CSDFT in the presence of an
external magnetic field.  In particular, we focus our attention  on
two-dimensional semiconductor quantum dots (QDs)~\cite{RM2002}  exposed to
uniform and constant external magnetic fields.  In addition to the various
applications in the field of  semiconductor nanotechnology, QDs are also
challenging test  cases for computational many-electron methods due to the
relatively large correlation effects. Moreover, the role of the current induced
by the external magnetic field is particularly relevant in QDs~\cite{elf}
making them a reference system in CDFT since its early
developements.~\cite{FV1994}
Therefore, it is interesting to examine whether the self-consistent solution of
CSDFT differs  from the result obtained by adding the external vector potential
to the SDFT scheme, which amounts to neglecting the xc vector potential of
CSDFT.

As expected, we find that  the bare exact-exchange (EXX) result is not
sufficient to obtain total energies in agreement with numerically accurate
quantum Monte Carlo (QMC) results, although a considerable improvement to the
Hartree-Fock result is found. However, including the self-interaction corrected
LSDA correlation energies to the EXX solution leads to  total energies that
agree very well with QMC results. In addition, within the given approximations,
our results  confirm that the role of self-consistent  calculations in the
framework of CSDFT is only minor. In particular,  we observe that accurate total
energies and densities can also be  obtained by simply modifying the SDFT scheme
by including the coupling to  the external vector potential. Indeed, this
procedure has been  employed in the past to partially remedy  the lack of good
approximate current-dependent functionals. Here, a validation is provided in the
more general context  of the OEP framework.

This paper is organized as follows. In Sec.~\ref{OEP}
we review the OEP method in CSDFT. The formalism is then adapted
to the case of QDs in magnetic fields in Sec.~\ref{application}.
In Sec.~\ref{remarks} we discuss details of the numerical 
procedure before presenting the results of our calculations
in Sec.~\ref{examples}. A brief summary is given in Sec.~\ref{summary}.

\section{Optimized effective potential method in CSDFT}
\subsection{General formalism}\label{OEP}

The Kohn-Sham (KS) equation in CSDFT reads (Har\-tree atomic units are used 
throughout unless stated otherwise)
\be
\left[\frac{1}{2}
\left(\!-i\nabla+\frac{1}{c} \v A_s(\v r)\right)^2\!\!\!
+v_s(\v r)+\mu_B \mbox{\boldmath{$\sigma$}} \v B_s(\v r)\right] \Phi_k = \varepsilon_k \Phi_k .
\label{ks-cdft}
\ee
The three KS potentials are given by 
\be
v_s(\v r)=v_0(\v r)+v_H(\v r)+ v_{xc}(\v r)+\frac{1}{2c^2}\left[\v A_0^2(\v r)-\v
A_s^2(\v r)\right],
\ee
\be
\v B_s(\v r) = \v B_0(\v r)+\v B_{xc}(\v r),
\ee
and
\be
\v A_s(\v r) = \v A_0(\v r)+\v A_{xc}(\v r),
\ee 
where the xc potentials are functional derivatives  of the
xc energy $E_{xc}$ with respect to the corresponding 
densities,
\bea
\label{vxc-c1}
v_{xc}(\v r) &=& \frac{\delta E_{xc}[n,\v m,\v j_p]}{\delta n(\v r)} \; ,
\eea
\bea
\label{bxc-c1}
\v B_{xc}(\v r) &=& - \frac{\delta E_{xc}[n,\v m,\v j_p]}{\delta \v m(\v r)},
\eea
and
\bea
\frac{1}{c} \v A_{xc}(\v r) &=& \frac{\delta E_{xc}[n,\v m,\v j_p]}
{\delta \v j_p(\v r)} \; ,
\label{axc-c1}
\eea
respectively.
The  self-consistency cycle is closed by calculating the density
\be
n(\v r) = \sum_{k=1}^{\mathrm{occ}} \Phi_k^{\dagger}(\v r) \Phi_k(\v r) \; ,
\ee
the magnetization density
\be
\v m(\v r) = - \mu_B \sum_{k=1}^{\mathrm{occ}} \Phi_k^{\dagger}(\v r) 
\mbox{\boldmath{$\sigma$}}
\Phi_k(\v r)  \; ,
\ee
and the paramagnetic current density 
\be
\v j_p(\v r) = \frac{1}{2 i}\! \sum_{k=1}^{\mathrm{occ}}\! \left[ \Phi_k^{\dagger}(\v r) 
\nabla \Phi_k(\v r) \!-\! (\nabla \Phi_k^{\dagger}(\v r) ) \Phi_k(\v r)\!
\right]\!.
\ee
The ground-state total energy of the interacting system can then be computed 
from
\bea\label{tot-enrg}
E[n,\v m,\v j_p] & = & T_s[n,\v m,\v j_p] + U[n] + E_{xc}[n,\v m,\v j_p]  \nonumber \\
& + & \int d\v r \; n(\v r) v_0(\v r)- \int d\v r\; \v m(\v r) \v B_0(\v r) \nonumber \\
& + & \frac{1}{c} \int d\v r\; \v j_p(\v r) \v A_0(\v r) \nonumber \\ 
& + & \frac{1}{2 c^2} \int d\v r \; n(\v r) \v A_0^2(\v r), 
\eea
where $T_s$  and $U$ are the kinetic energy of the KS system and the 
Hartree energy, respectively.

Gauge invariance of the energy functional implies that $E_{xc}$ depends on 
the current only through the vorticity,  
\be
\mbox{\boldmath{$\nu$}}(\v r) = \nabla \times (\v j_p(\v
r)/n(\v r) ), 
\ee
i.e., 
$E_{xc}[n,\v j_p, \v m] = \bar{E}_{xc}[n,\mbox{\boldmath{$\nu$}},\v m]$.~\cite{VR1988}
This immediately leads to the following relation for the xc
vector potential
\be
\nabla \left( n(\v r) \v A_{xc}(\v r) \right) = 0 \; .
\label{gauge-prop}
\ee

If one uses an approximate $E_{xc}$ which is given \emph{explicitly} in terms of
the densities, the calculation of the corresponding xc potentials via Eqs.
(\ref{vxc-c1})-(\ref{axc-c1}) is straightforward. Here, however, we deal with
approximations to the xc energy which are  explicit
functionals of the KS spinor orbitals $\Phi_k$. These  functionals are,
via the Hohenberg-Kohn theorem, {\em implicit} functionals of  the densities. In
the spirit of the original OEP formalism, the corresponding
integral equations for the xc potentials can be
derived~\cite{PKHG2006} by requiring that
the effective fields minimize the value of the ground-state  total energy
(\ref{tot-enrg}). Therefore, the functional derivatives of the total
energy with respect to the three KS potentials are required to vanish. 
This procedure leads to three OEP equations  which are most
conveniently written as \cite{PKHG2006}  
\be
\label{oep1}
\sum_{k=1}^{\mathrm{occ}} \Phi_k^{\dagger}(\v r)\Psi_k(\v r) + h.c. = 0, \; 
\ee
\be
\label{oep2}
-\mu_B\sum_{k=1}^{\mathrm{occ}} \Phi_k^{\dagger}(\v r)\mbox{\boldmath{$\sigma$}}
\Psi_k(\v r) + h.c. 
= 0 \;,
\ee
and
\be
\label{oep3}
\frac{1}{2i}\sum_{k=1}^{\mathrm{occ}} 
\left[\Phi_k^{\dagger}(\v r)\nabla\Psi_k(\v r)
-\left(\nabla\Phi_k^{\dagger}(\v r)\right)\Psi_k(\v r)\right] -
h.c. = 0 \; ,
\ee
where we have defined the so-called ``orbital shifts'' 
\cite{GKKG2000,KP2003a}
\be
\Psi_k(\v r)=\sum_{\stackrel{\scriptstyle j=1}{j\neq
k}}^{\infty}\frac{D_{kj}^{\dagger}\Phi_j(\v r)}{\epsilon_k-\epsilon_j}\;,
\label{orb-shift}
\ee
with
\begin{multline}
D_{kj}^{\dagger}= \int d\v r' \biggl\{
v_{xc}(\v r')\Phi_j^{\dagger}(\v r')\Phi_k(\v r') \\
+\frac{1}{2 i c}\v A_{xc}(\v r)
\left[\Phi_j^{\dagger}(\v r')\nabla'\Phi_k(\v r')
-\left(\nabla'\Phi_j^{\dagger}(\v r')\right)\Phi_k(\v r')\right] \\
+\mu_B\v B_{xc}(\v r')\Phi_j^{\dagger}(\v r')\mbox{\boldmath{$\sigma$}}
\Phi_k(\v r') -\Phi_j^{\dagger}(\v r')\frac{\delta E_{xc}}{\delta 
\Phi_k^{\dagger}(\v r')} \biggl\}.
\end{multline}
The orbital shifts $\Psi_k$ have the structure of a first-order shift from  the
unperturbed orbital $\Phi_k$ under a perturbation whose matrix elements  are
given by $D_{kj}^{\dagger}$. Physically, the OEP equations
(\ref{oep1})-(\ref{oep3}) then imply that the densities do not change  under
this perturbation. If $\v A_{xc}$ is set to zero, Eqs.~(\ref{oep1}) and
(\ref{oep2}) reduce exactly to the OEP equations of non-collinear SDFT.
\cite{SDDHKGSN2006}

Eqs.~(\ref{oep1}) - (\ref{oep3}) form a set of coupled integral equations for the
three unknown xc potentials, and they  can be solved by a direct computation of
the orbital shifts.\cite{KP2003a, SDDHKGSN2006} Alternatively, one can employ
the Krieger-Li-Iafrate (KLI) approach as a simplifying approximation
\cite{SH1953,KLI1992b} which is known to yield potentials which are very close
to the full OEP ones in SDFT. In the following we utilize the KLI approximation
in the description of a quasi-two-dimensional semiconductor  QD~\cite{RM2002} in
an external magnetic field. 

\subsection{Application to quantum dots}\label{application}

The QD is described as a many-electron system restricted to the $xy$ plane and
confined in that plane by an external parabolic potential  $v_0=\frac{1}{2} m^*
\omega_0^2r^2$ with $r^2=x^2+y^2$. Following the most common
experimental setup,~\cite{RM2002} the external magnetic field is defined to be
uniform and perpendicular to the $xy$ plane, i.e., $\v B_0(\v r) = \nabla \times
\v A_0(\v r)=B_0\v e_z$ with the gauge $\v A_0(\v r)=B_0 r \v e_\theta/2$. We apply
the effective-mass approximation with the material parameters for GaAs, i.e.,
the effective mass $m^*=0.067$, the dielectric constant, $\epsilon^*=12.4$, and
the effective gyromagnetic ratio $g^*=-0.44$. 

In QDs the magnetization is parallel to the external field, i.e.,
these systems show collinear magnetism.
Therefore, the KS magnetic field $\v B_s$ and the magnetization density
have only non-vanishing $z$-components.  The Pauli-type KS equation becomes
diagonal in spin space and can be decoupled into two separate equations for the
spin-up and spin-down orbitals $\varphi_{k\sigma}(\v r)$. We further assume that
the xc potentials preserve the cylindrical symmetry of the problem, i.e., 
\be
\label{symmvxc}
v_{xc\sigma}(\v r) = v_{xc\sigma}(r) = v_{xc}(r) \pm 
\mu_B g^* B_{xcz}(r), 
\ee
where the upper signs are for spin-up and lower signs for spin-down electrons, and $
\v A_{xc}(\v r) = A_{xc}(r)\v e_{\theta}$. Due to the cylindrical symmetry we can
separate the wave function into radial and angular parts as
$\varphi_{jl\sigma}(\v r)=\exp(il\theta)R_{jl\sigma}(r)$, where the radial wave
functions $R_{jl\sigma}(r)$ are real-valued eigenfunctions of the 
Hamiltonian
\bea\nonumber
\lefteqn{\hat{H}_{sl\sigma}=
-\frac{1}{2m^*}\left(\frac{\partial^2}{\partial r^2}
+\frac{1}{r}\frac{\partial}{\partial r} -\frac{l^2}{r^2}\right)
+\frac{l}{2}\omega_c +m^*\frac{\Omega^2}{2}r^2}\\
&& +\frac{l}{m^*c}\frac{A_{xc}(r)}{r} 
\pm\mu_B m^* g^* B_0 +v_H(r)+v_{xc\sigma}(r)
\label{hsradial}
\eea
with the total confinement $\Omega=\sqrt{\omega_0^2+\omega_c^2/4}$, and the cyclotron frequency 
$\omega_c=B_0/m^*c$.  The radial wave functions are expanded
in the basis of eigenfunctions of the corresponding non-interacting problem,
i.e., the eigenfunctions of the 
Hamiltonian (\ref{hsradial}) with the Hartree and
all xc potentials set to zero.

As a consequence of the cylindrical symmetry, 
the densities are independent of the angle and
thus given solely in terms of $r=|\v r|$. Also, only the
$\theta$-component of the paramagnetic current density, as the conjugate
variable to the vector field in this direction, plays a role, i.e.,  
$\v j_p(\v r)=(j_{p\uparrow}(r)+j_{p\downarrow}(r))\v e_\theta$. 
Instead of using the density and the $z$-component of the
magnetization, one employs the spin-up and
spin-down densities. Hence, the three densities to be determined are
$n_\uparrow(r)$, $n_\downarrow(r)$, and $j_p(r)$. 

Consequently, the OEP-KLI
equations are given as a $3\times 3$ matrix equation which reads
\be\label{symmoep}
{\cal D}(r){\cal V}_{xc}(r)={\cal R}(r),
\ee
where the potential vector is given by
\be
{\cal V}_{xc}(r)=\left(v_{xc\uparrow}(r),v_{xc\downarrow}(r), 
\frac{1}{c}A_{xc}(r)\right).
\ee
The matrix $\cal D$ reads
\be\label{symmcald}
{\cal D}=\left(
\begin{array}{ccc}
n_{\uparrow}(r) & 0 & j_{p\uparrow}(r)\\
0 & n_{\downarrow}(r) & j_{p\downarrow}(r)\\
j_{p\uparrow}(r) & j_{p\downarrow}(r) & N(r)
\end{array}\right),
\ee
where the densities and current densities are given by
\bea
n_{\sigma}(r)=\sum_{\{jl\}}^{\mathrm{occ}} R_{jl\sigma}^2(r),\\
j_{p\sigma}(r)=\sum_{\{jl\}}^{\mathrm{occ}}\frac{l}{r}R_{jl\sigma}^2(r).
\eea
The last component $N(r)$ in Eq.~(\ref{symmoep}) reads
\be\label{N}
N(r)=\sum_{\sigma=\uparrow,\downarrow}\sum_{\{jl\}}^{\mathrm{occ}}
\frac{l^2}{r^2}R_{jl\sigma}^2(r).
\ee

The right-hand-side of Eq. (\ref{symmoep}) contains functional 
derivatives of the xc energy. They can be calculated once an 
approximation to the xc energy is specified. Here, we use the 
EXX approximation to $E_{xc}$, i.e.,
\bea
\lefteqn{E_x^{\mathrm{EXX}} = 
-\frac{1}{2}\sum_{\sigma=\uparrow,\downarrow}
\sum_{\{jl\},\{km\}}^{\mathrm{occ}}} \nonumber \\
\label{exxcoll}
&&\int d^2r \, d^2r'\; \frac{\varphi_{jl\sigma}^*(\v r')\varphi_{jl\sigma}(\v r)
\varphi_{km\sigma}(\v r')\varphi_{km\sigma}^*(\v r)}
{\epsilon^*\mid\v r-\v r'\mid} \; .
\eea
The first two components of $\cal R$ on the RHS of 
Eq. (\ref{symmoep}) are then given by
\bea\nonumber
{\cal R}_{1,2}(r)&=&-\frac{1}{2} 
\sum_{\{jl\},\{km\}}^{\mathrm{occ}}
R_{jl\sigma}(r)R_{km\sigma}(r) \\
&\times & \int d^2r' \frac{e^{i\theta'(l-m)}R_{jl\sigma}(r')R_{km\sigma}(r')}
{\sqrt{r^2+r'^2-2rr'\cos{\theta'}}}\nonumber\\
&-&\frac{1}{2} \sum_{\{jl\}}^{\mathrm{occ}} 
n_{jl\sigma}(r)D_{jl,jl,\sigma}^*+c.c.,
\eea
where for ${\cal R}_1$ $\sigma=\uparrow$ and for 
${\cal R}_2$ $\sigma=\downarrow$. The third component is given by
\bea\nonumber
{\cal R}_3(r)&=&-\frac{1}{2}\sum_{\sigma=\uparrow,\downarrow}
\sum_{\{jl\},\{km\}}^{\mathrm{occ}}
\frac{l+m}{2}R_{jl\sigma}(r)R_{km\sigma}(r) \\
&\times & \int d^2r' \frac{e^{i\theta'(l-m)}R_{jl\sigma}(r')R_{km\sigma}(r')}
{\sqrt{r^2+r'^2-2rr'\cos{\theta'}}}\nonumber\\
&-&\frac{1}{2}\sum_{\sigma=\uparrow,\downarrow}
\sum_{\{jl\}}^{\mathrm{occ}}
j_{pjl\sigma}(r)D_{jl,jl,\sigma}^*+c.c.
\eea
with
\bea\nonumber
D_{jl,jl\sigma}^*&=&\int\! d^2r\: 
\left(v_{xc\sigma}(r)+\frac{l}{c}A_{xc}(r)\right)R_{jl\sigma}^2(r)\\
\nonumber
&+&\sum_{\{km\}}^{\mathrm{occ}}\int\!\!\int d^2r d^2r' e^{i\theta'(l-m)}\\
&\times&\frac{R_{jl\sigma}(r')R_{km\sigma}(r')
R_{jl\sigma}(r)R_{km\sigma}(r)}
{\sqrt{r^2+r'^2-2rr'\cos{\theta'}}}
\eea
in all three cases.

\section{Numerical Results}\label{numerical}

\subsection{General remarks}\label{remarks}

A detailed analysis of Eq.~(\ref{symmoep}) reveals  that for a system with a
vanishing current the  third line of the matrix equation vanishes identically.
However, for these states the correct value of the current is already obtained
at the level of  SDFT as a natural symmetry constraint. In fact, using zero 
vector potential as the initial value, one  can show that it remains zero at
each iteration.  Hence, one recovers the original SDFT result for non-current
carrying states.~\cite{PKHG2006}  On the other hand, for current-carrying states
the xc vector  potential is always non-vanishing even if one chooses a 
vanishing vector potential as the initial value.

A closer inspection of the KLI equations  shows that they become linearly
dependent in the asymptotic  region and therefore do not have a unique
solution.  In our numerical procedure, we take a pragmatic approach to the 
problem of linearly dependent KLI equations  and add a very small positive
constant to $N(r)$ in Eq.~(\ref{N}). As the consequence, the limit becomes 
$A_{xc}(r) \stackrel{ r \to \infty}{\longrightarrow} 0$. In addition, we impose 
$v_{xc,\sigma}(r) \stackrel{ r \to \infty}{\longrightarrow} -1/r$.  This
procedure also limits the possible  appearance of numerical artifacts in the KLI
potentials resulting from a finite basis-set. Such difficulties have also
occurred for open-shell  atoms.~\cite{PKHG2006,Diemo} Although we face similar
problems in QD calculations (see below), we have confirmed that the evaluation
of the total energies, densities, and currents is not considerably affected. A
further analysis is presented elsewhere.~\cite{PhD} In the context of the full
solution of the OEP equations, problems in the   computation of the effection
potential due to the use of a finite basis-set have been recently analyzed in
several works, and different possible solutions have been
proposed.~\cite{TimH,Staroverov1,Staroverov2,RohrGri,Hesselmann}

\subsection{Examples}\label{examples}

\begin{figure}
\includegraphics[width=0.85\columnwidth]{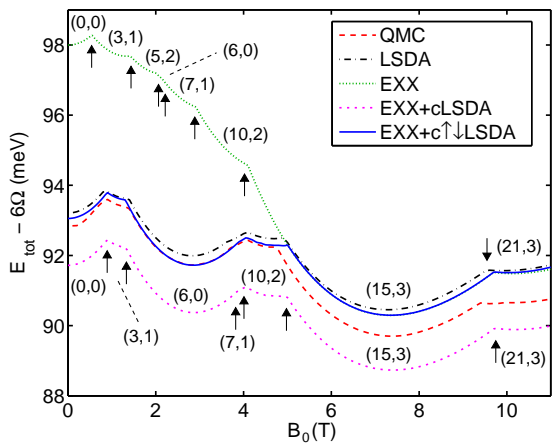}
\caption{\label{totalenergy} 
(color online). Total ground-state energy
(minus $6\Omega=6\sqrt{\omega_0^2+\omega_c^2/4}$) in 
a six-electron quantum dot as a function of external
magnetic field (SI units). The results have been 
calculated using the exact-exchange (EXX), 
EXX with LSDA correlation 
(EXX+cLSDA), and EXX with the corrected LSDA 
correlation (EXX+c$\uparrow\downarrow$LSDA). 
The LSDA and quantum Monte Carlo (QMC)
results \cite{SRSHPN2003} are shown for comparison.
The arrows mark the points where the ground-state
configuration ($L_z,S_z$) changes.
}
\end{figure}

Figure~\ref{totalenergy} 
shows the total energy 
of a six-electron QD ($\omega_0=5$~meV) as a 
function of $B_0$. The kinks correspond to changes 
in the ground-state configuration ($L_z,S_z$). Apart 
from the fully-polarized ($S_z=3$) states, the EXX 
energies (dotted line) are considerably too large when compared with
the accurate QMC results (dashed line).~\cite{SRSHPN2003} 
EXX also leads to an erroneous occurrence of the ($-5,2$) 
ground-state at $B_0=1.5...2.0$~T. However, adding
the LSDA correlation~\cite{AMGB2002} post-hoc 
to the EXX energies (EXX+cLSDA) yields the correct sequence of 
states as a function of $B_0$. 
This is a major improvement over the cLSDA-corrected
Hartree-Fock calculation which does not give the correct
ground states for a similar system.~\cite{X2000}
As expected, the corrections given by cLSDA are
largest for the unpolarized state ($0,0$) and smallest 
for the completely polarized states ($-15,3$) and 
($-21;3$). This is due to the fact that the electron exchange 
has a larger effect on the total energy
in systems with a high number of same-spin electrons.

Despite the improvement of EXX+cLSDA over the bare EXX, 
the result is not satisfactory in comparison with QMC:
Figure~\ref{totalenergy} shows that the energies of 
EXX+cLSDA are consistently too low by $1.0-1.5$~meV. 
On the other hand, the agreement between QMC and the 
{\em conventional} LSDA (dash-dotted line) is very 
good. Hence, taking into account that the EXX is expected to 
capture the {\em true} exchange energy by a good accuracy (the only 
deviation arising from the missing correlation in the 
self-consistent solution), our result demonstrates
the inherent tendency of the LSDA to cancel out its respective 
errors in exchange and correlation. This well-known error
cancellation is lost when adding LSDA correlation to the 
EXX result. As expected, the performance of EXX+cLSDA with respect to QMC is at its
best in the fully polarized regime ($B_0 \gtrsim 5$ T), where the exchange
contribution in the total energy is relatively at largest.

As a simple cure to the error in EXX+cLSDA,
we apply a type of self-interaction correction as
first suggested by Stoll and co-workers.~\cite{stoll}
The LSDA correlation energy can be improved by
\bea\nonumber
\lefteqn{
E_{\rm c{\uparrow}{\downarrow}LSDA}=E_{\rm cLSDA}}\hspace{0.5cm}\\
&-&\!\int d^2 r \,
\bigl\{ n_{\uparrow}(\v r)\epsilon_c[n_\uparrow,0]
+n_{\downarrow}(\v r)\epsilon_c[0,n_\downarrow]\bigl\}\!,
\eea
where $\epsilon_c[n_\uparrow,n_\downarrow]$ is the correlation energy
per electron in the two-dimensional electron gas.~\cite{AMGB2002}
Therefore, in this approximation, denoted as
EXX+c${\uparrow}{\downarrow}$LSDA, the correlation energy between
like-spin electrons is removed. We emphasize that
this contribution is non-zero in the {\em exact} treatment and
thus cannot be neglected. However, within the LSDA it contains
mostly self-interaction energy. Now, we find that 
EXX+c${\uparrow}{\downarrow}$LSDA (solid line) is very
close to QMC, and actually performs better than the conventional 
LSDA. 

\begin{figure}
\includegraphics[width=0.92\columnwidth]{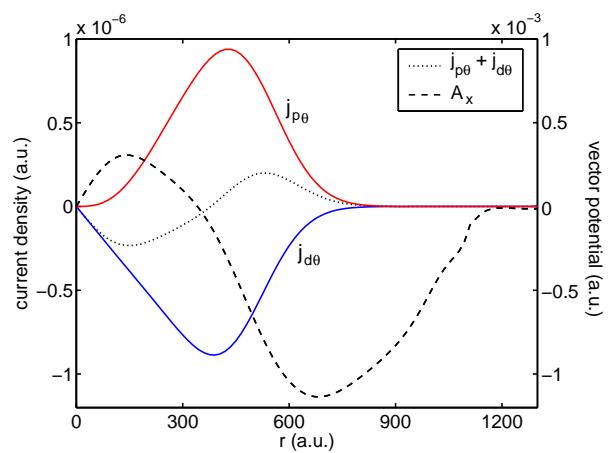}
\caption{\label{currents} (color online). Angular components 
of the paramagnetic and diamagnetic
currents, $j_{p\theta}$ and $j_{d\theta}$, and 
their sum for the ($-21,3$) state at $B_0=11$~T. 
The dashed line shows the exchange vector potential 
for the same configuration.}
\end{figure}

Figure~\ref{currents} 
shows the paramagnetic current 
$j_{p\theta}$ and the diamagnetic current 
$j_{d\theta}(r)=n(r)A_{0\theta}(r)/m^*$ at $B_0=11$~T 
for the ($-21,3$) state. The total current 
$j_\theta=j_{p\theta}+j_{d\theta}$ changes sign at 
$r\sim 350$~a.u. due to the existence of a single vortex 
at the center of the QD. We find the vortex solution 
in agreement with both LSDA and numerically exact
calculations.~\cite{SHPN2004} 

In Fig.~\ref{currents} we also show the exchange vector  potential $A_x$.  The
small kink at $r\sim 1100$~a.u. is due to a basis-set problem described in
Sec.~\ref{remarks}. The maximum of $|A_x|$ is located near the edge  of the QD
at $r\sim 700$~a.u. However, its relative magnitude with respect to the external
vector  potential $A_0$ is largest at $r\sim 150$ a.u., where we find
$|A_x/A_0|\sim 0.1$. Despite the considerable  magnitude of $A_x$, we find that
its effect on physical quantities like the total energy, density, and  current
density is practically negligible. In the case presented in Fig.~\ref{currents},
for example, the difference between SDFT and CSDFT total energies  is $\sim
0.02\%$. In the context of  the OEP method, the minor role of the xc vector
potential has been observed for  open-shell atoms, \cite{PKHG2006} molecules,
\cite{LHC1995} and extended systems.~\cite{SPKSDG07}  Earlier QD studies in the
level of LSDA have also led to similar conclusions.~\cite{SRSHPN2003}

Finally, we point out that, in principle, a given functional 
should be evaluated with KS orbitals obtained from self-consistent 
calculations and not in a post-hoc manner as we have done
in this work.
However, the variational nature of DFT implies that 
if one evaluates the total energy with a density which
slightly differs from the self-consistent density, the resulting 
change in the energy is of second order in the small deviation of 
the densities. 

\section{Summary}\label{summary}

We have applied the optimized effective potential method in 
current-spin-density functional theory to two-dimensional 
systems exposed to external magnetic fields.
We have observed that the bare exact-exchange result
(within the KLI approximation) is not sufficient in finding the 
correct ground-state sequence as a function of the
magnetic field, although a considerable improvement over 
the Hartree-Fock results is found. 
Adding the correlation energy in the form of the standard 
local spin-density approximation yields excellent agreement 
of the ground-state energies with quantum Monte Carlo results, if  
the spurious self-interaction error is corrected.
Moreover, within the specified approximations, we found no  
considerable differences in total energies and densities
when comparing the results obtained using a full-fledged 
current-spin-density functional theory and a spin-density functional 
scheme modified to include the coupling to the external vector potential.

\begin{acknowledgments} 

The authors like to thank H. Saarikoski and A. Harju for providing us  with
their data published in Fig.~3 of Ref.~\onlinecite{SRSHPN2003}. We are grateful
to Diemo K\"odderitzsch for the valuable discussions on the properties of the
effective potentials and the issue of numerical artifacts in the optimized
potentials. We gratefully acknowledge the financial support through the
Deutsche  Forschungsgemeinschaft, the EU's Sixth Framework Program through  the
Nanoquanta Network of Excellence (NMP4-CT-2004-500198), the Academy of Finland,
and the Finnish Academy of  Science and Letters through the Viljo, Yrj{\"o} and
Kalle  V{\"a}is{\"a}l{\"a} Foundation.

\end{acknowledgments}

\end{document}